\documentstyle[12pt]{article}

\newcommand{\beq}{\begin{equation}}
\newcommand{\eeq}{\end{equation}}
\newcommand{\beqa}{\begin{eqnarray}}
\newcommand{\eeqa}{\end{eqnarray}}
\newcommand{\beqar}{\begin{eqnarray*}}
\newcommand{\eeqar}{\end{eqnarray*}}
\renewcommand{\a}{\alpha}
\renewcommand{\b}{\beta}

\newcommand{\cH}{{\cal H}}

\newcommand{\g}{\gamma}

\renewcommand{\L}{\Lambda}

\renewcommand{\O}{\Omega}
\newcommand{\p}{\phi}

\renewcommand{\t}{\theta}

\newcommand{\m}{\mu} 
\newcommand{\A}{{\cal A}}
\newcommand{\ssc}{\scriptscriptstyle}
\newcommand{\ssB}{{\ssc B}}
\newcommand{\ssR}{{\ssc R}}

\newcommand{\ie}{{\it i.e.,}\ }
\newcommand{\eqr}[1]{eq.~(\ref{#1})\ }
\newcommand{\eqc}[1]{eq.~(\ref{#1})}

\newcommand{\rrc}[1]{(\ref{#1})}

\newcommand{ \hrz }{r_+}
\newcommand{\prmrp}{\left (  1-\frac{r_-}{r} \right )
 \left (  1- \frac{r_+}{r} \right )}
\newcommand{\prmrpb}{ \left [    \left(  1-\frac{r_-}{r}   \right)
   \left( 1- \frac{r_+}{r} \right) \right ]}
\newcommand{\RN}{Reissner-Nordstrom}

\begin{document}
\begin{titlepage}
\rightline{\small gr-qc/9503003 \hfill McGill/95-06}
\vskip 5em

\begin{center}
{\bf \huge Black hole entropy without brick walls}
  \vskip 3em

{\large
	Jean-Guy Demers\footnote{jgdemers@hep.physics.mcgill.ca},
 	Ren\'e Lafrance\footnote{lafrance@hep.physics.mcgill.ca}
and  	Robert C. Myers\footnote{rcm@hep.physics.mcgill.ca} \\[1.5em] }
\em{	Department of Physics, McGill University \\
	Montr\'eal, Qu\'ebec, Canada H3A 2T8}

\vskip 4em

\begin{abstract}
We present evidence which confirms
a suggestion by Susskind and Uglum regarding black hole entropy.
Using a Pauli-Villars regulator, we find that
't Hooft's approach to evaluating black hole entropy through
a statistical-mechanical counting of states for a scalar field
propagating outside the event horizon
yields precisely the one-loop renormalization of the standard
Bekenstein-Hawking formula, $S=\A/(4G)$. Our calculation also yields
a constant contribution to the black hole entropy, a contribution associated
with the one-loop renormalization of higher curvature terms in the
gravitational action.
\end{abstract}
\end{center}

\vskip 3em
\end{titlepage}

\section{Introduction}

It is now over twenty years since Bekenstein introduced the idea
that black holes carry an intrinsic entropy proportional
to the surface area of the event horizon measured in
Planck units, \ie $\A/{\ell_p}^2$ \cite{bekenstein}.
Hawking's discovery\cite{hawking} that, in quantum field
theory, black holes actually
generate thermal radiation allowed the determination of a precise
formula for this entropy\footnote{
We adopt the standard conventions of setting $\hbar=c=
k_{\scriptscriptstyle B}=1$, but we will explicitly retain
Newton's constant, $G$, in our analysis.
Also, we will employ the metric and curvature conventions of
Ref.~\cite{waldtext}.}: $S={\A/(4G)}=\A/(4{\ell_p}^2)$.
This Bekenstein-Hawking formula is applicable for any black hole
solution of Einstein's equations.
Recently, it was shown[4--8] 
that when gravity is described by a higher-curvature
effective action, the Bekenstein-Hawking result
is only the leading contribution in
an integral of an entropy density over a cross-section of the
horizon, \ie $S=\oint_\cH d^2x\sqrt{h}\,\rho_s$. The
contribution of the Einstein-Hilbert action to the entropy density
is simply the constant $1/(4G)$, which then yields the expected
result $\A/(4G)$. Any higher curvature interactions make additional
higher curvature contributions to $\rho_s$.

Our understanding of black hole entropy, though, is only within
a thermodynamic framework, and despite a great deal of effort,
a microphysical understanding of this entropy is still lacking.
Many attempts have been made to provide a definition of black
hole entropy using statistical mechanics. York \cite{york}
suggested that the entropy be considered as the
logarithm of the number of ways that the ``quantum ergosphere'' can be
excited during the evaporation of a Schwarzschild black hole
into a surrounding thermal bath. This model has the unsatisfying feature
of being nonlocal in time since this entropy includes contributions from
the entire future evolution of the black hole.
Within the membrane paradigm\cite{zurek}, the entropy is associated
with the thermal bath of quantum fields perceived by stationary
observers under the stretched horizon. In a related approach
introduced by 't Hooft\cite{thooft}, the entropy arises from a
thermal bath of fields propagating just outside the horizon
(see also Ref.~\cite{mann}).
Recently, there has been a great deal of interest in an interpretation
of black hole entropy as entanglement entropy. One defines a density
matrix $\rho$ by starting with the vacuum state of some quantum
field, and tracing over the field degrees
of freedom inside the horizon\cite{sorkin}.
The entropy is then given by the standard formula
$S=-{\rm Tr}(\rho \, \log\rho)$. Alternatively, Frolov and Novikov
suggest that one should trace out the degrees of freedom external
to the horizon\cite{frolova}. Both of the latter approaches should
yield the same result as long as the initial global state of the
quantum field is a pure state\cite{srednicki}. Further, Kabat and
Strassler argued that the density operator constructed with such
a trace has a thermal character
independent of the details of the quantum field theory\cite{kabat}.
The latter result draws a connection between the entanglement
entropy analysis and the two previous approaches.

Another feature common to all four of these calculations is that
they yield a black hole entropy proportional to the
surface area, but with a divergent coefficient. Thus one must introduce
a cut-off to regulate any of these results. For example, 't Hooft introduces
a ``brick wall'', a fixed boundary near the horizon within which
the quantum field does not propagate.
Susskind and Uglum suggested that these divergences have the correct form to
be absorbed in the Bekenstein-Hawking formula as a renormalization
of Newton's constant\cite{susskind}. Thus these calculations
should be regarded as yielding the one-loop correction of
quantum field theory to the black hole entropy\cite{callan,tedb}.
Further these authors suggested that the ``bare entropy'' may
have a sensible statistical mechanical interpretation in the context
of string theory\cite{susskind}. The latter conjecture has
generated a great deal of interest in understanding black hole
entropy in the context of string theory\cite{string}.

The purpose of this paper is to investigate the first conjecture
of Susskind and Uglum connecting entropy divergences with the
renormalization of the coupling constants in the theory. In particular,
we test this conjecture
by examining a scalar field propagating in a
four-dimensional, non-extremal Reissner-Nordstrom
black hole background.
We begin in section \ref{effsect}
by considering the renormalization of the coupling constants in the
gravitational action by a quantum scalar field theory.
We regulate the scalar field
loops using a Pauli-Villars scheme, and determine the precise
renormalization of  Newton's constant. In section \ref{entropysect},
we present 't Hooft's approach to calculating the black hole entropy.
The advantage of the Pauli-Villars regulator is evident at this stage
since it can also be used to implement a cut-off for the entropy calculation.
Thus we can remove 't Hooft's brick wall (\ie the explicit length
cut-off in Ref.~\cite{thooft}) and we can compare the results with those
found for the effective action. The regulated entropy takes
the form: $S=B'\,\A/4  + A'$,
where $A'$ and $B'$ are constants which have quadratic and logarithmic
dependences on the Pauli-Villars mass respectively.
In the final section, we compare
the results of the two previous calculations.
We find that $B'$ is precisely the same constant found in the
renormalization of Newton's constant, while $A'$ is related to the
renormalization of certain higher-curvature interactions. We conclude
with a discussion of these results, including a comment on   the
extremal \RN\ black hole.

\section{Renormalization of the gravitational action } \label{effsect}
In the study of the one-loop effective action\cite{birrell},
one may start with the gravitational action
\beq
I_g=\int d^4x\, \sqrt{-g} \left[ -\frac{\L_\ssB}{8\pi G_\ssB} +
\frac{R}{16\pi G_\ssB} +\frac{\a_\ssB}{4\pi} R^2 +
\frac{\b_\ssB}{4\pi} R_{ab}R^{ab} +
\frac{\g_\ssB}{4\pi} R_{abcd}R^{abcd} + \ldots\ \right]
\label{actg}
\eeq
where $\L_\ssB$ is the cosmological constant, $G_\ssB$ is Newton's constant,
while $\a_\ssB$, $\b_\ssB$ and $\g_\ssB$ are dimensionless coupling
constants for the interactions which are quadratic in the curvature.
The subscript $B$ indicates that all of these constants are the bare coupling
constants. The ellipsis indicates
that the action may also include other covariant
higher derivative interactions, but only those terms explicitly shown
will be of interest in the present analysis.
We also include the action for a minimally coupled neutral scalar field,
\beq
I_m=-\frac{1}{2} \int d^4x \, \sqrt{-g} \left[ g^{a b} \nabla_{\! a}\phi
 \nabla_{\! b} \phi +m^2 \phi^2 \right]\ .
\label{miniaction}
\eeq
Here we wish to determine the effective action for the metric which
results when in the path integral the scalar field is integrated out.
In the present case, this integration is simply gaussian, yielding
the square root of the determinant of the propagator; the
contribution to the effective gravitational
action, which is essentially the logarithm of this result,
is then
given by\cite{birrell}: $W(g)=-{i\over2}{\rm Tr}\log[-G_{\ssc F}(g,m^2)]$.
Of course, as it stands, this expression is divergent and must be
regulated to be properly defined. The divergences of this one-loop
effective action,
as well as its metric dependence, are easily identified using
an adiabatic expansion for the DeWitt-Schwinger proper time representation
of the propagator\cite{dewitt}.
This leads to a representation of the scalar one-loop action as
an asymptotic series\cite{gilk}:
\beq
W(g)=-{1\over32\pi^2}\int d^4x\,\sqrt{-g}\,
\int_0^\infty{ds\over s^3}\sum_{n=0}^\infty
a_n(x)\,(is)^n e^{-im^2s}
\label{leff}
\eeq
where the adiabatic expansion coefficients $a_n(x)$ are functionals of
the local geometry at $x$. Thus, they are local expressions constructed
in terms of the metric and the curvature tensor. For example,
\beqa
a_0(x)&=&1 \nonumber\\
a_1(x)&=&\frac{1}{6} R \nonumber\\
a_2(x) &=& \frac{1}{180} R^{abcd}R_{abcd} -\frac{1}{180}
 R^{ab}R_{ab} +\frac{1}{30} \Box R +\frac{1}{72}R^2\ .
\label{expcoeff}
\eeqa
In the present case of four dimensions,
the ultraviolet divergences arise as
$s\rightarrow0$ in the first three terms of the series \rrc{leff}.

The effective action may be regulated using many different
methods\cite{birrell}, but in the present calculation we adopt
a Pauli-Villars regularization scheme\cite{pvreg}. In general,
such a scheme involves the introduction of a number
of fictitious fields with very
large masses set by some ultraviolet cut-off scale. Some
of these regulator fields are also quantized with the ``wrong'' statistics,
so that their contributions in loops tend to cancel those of
the remaining fields. The number, statistics and masses of the
regulator fields are chosen in order to render all of the
ultraviolet divergences finite. In the present four-dimensional scalar
field theory, one introduces five regulator fields: $\phi_1$ and $\phi_2$,
which are two anticommuting fields with mass $m_{\ssc 1,2}=\sqrt{\m^2+m^2}$\ ;
$\phi_3$ and $\phi_4$, which are two commuting fields with mass $m_{\ssc 3,4}=
\sqrt{3\m^2+m^2}$\ ; and $\phi_5$, which is an anticommuting
field with mass $m_{\ssc 5}=\sqrt{4\m^2+m^2}$.
The total action for the matter fields then becomes
\beq
I_m=-\frac{1}{2}\sum_{i=0}^5
 \int d^4x \, \sqrt{-g} \left[ g^{a b} \nabla_a \phi_i
 \nabla_b \phi_i +m_i^2 \phi_i^2 \right]\
\label{pvaction}
\eeq
where the original scalar is included as $\phi_{\ssc 0}=\phi$ with
mass $m_{\ssc 0}=m$. Now, each field makes a contribution
to the effective action as discussed above, except that as a result of
the anticommuting statistics for $\phi_2,\ \phi_3$ and $\phi_5$,
their contribution to the effective action  has the opposite sign,
\ie $W(g)\simeq+{i\over2}{\rm Tr}\log[-G_{\ssc F}(g,m_i^2)]$.
Let us focus on the divergent terms in eq.~\rrc{leff}, since
these are the ones for which the regulator fields
make significant contributions, we then obtain
\beqa
W_{div}&=& -\frac{1}{32\pi^2}\int d^4x\,\sqrt{-g}\,
 \int_0^\infty \frac{ds}{s^3} \,
 \left[ a_0(x) +isa_1(x) +(is)^2 a_2(x) \right]  \nonumber \\
&  & \qquad\qquad\times\ \left[ e^{-im^2s}
-2e^{-i(\m^2+m^2)s}+2e^{-i(3\m^2+m^2)s} -e^{-i(4\m^2+m^2)s} \right]
\nonumber \\
&=&\frac{1}{32\pi^2}\int d^4x\,\sqrt{-g}\, \left[ -C\,a_0(x)+B\, a_1(x) +
A a_2(x) \right]\ .
\label{diver}
\eeqa
In this expression, $A,B$ and $C$ are constants
which depend on $m$ and $\m$, and which diverge for
$\m\rightarrow \infty$:
\beqa
A&=& \ln \frac{4\m^2+m^2}{m^2} +2\ln \frac{\m^2+m^2}{3\m^2+m^2}
 \label{Acoeff} \\
B&=& \m^2 \left[2 \ln \frac{3\m^2+m^2}{\m^2+m^2}
+4\ln \frac{3\m^2+m^2}{4\m^2+m^2} \right] +
m^2 \left[ \ln \frac{m^2}{4\m^2+m^2} +2\ln \frac{3\m^2+m^2}
{\m^2+m^2} \right] \label{Bcoeff} \\
C&=&   \m^4 \left[
8\ln \frac{3\m^2+m^2}{4\m^2+m^2} +\ln \frac{3\m^2+m^2}
{\m^2+m^2}\right]  +2  m^2\m^2 \left[  \ln \frac{3\m^2+m^2}
{\m^2+m^2} +2\ln \frac{3\m^2+m^2}{4\m^2+m^2} \right] \nonumber \\
& &\qquad +\frac{m^4}{2} \left[ \ln \frac{m^2}{4\m^2+m^2} +2\ln
\frac{3\m^2+m^2}{\m^2+m^2} \right]\ \ . \nonumber
\eeqa
Combining the scalar one-loop action with the original bare action in
eq.~\rrc{actg}, we can identify the renormalized coupling
constants in the effective gravitational action
\beqa
\lefteqn{I_{eff}
=I_g\ +\ W}
\nonumber \\
&& =\int d^4 x \sqrt{-g} \left[   -\frac{1}{8\pi} \left( \frac{\L_\ssB}{G_\ssB}
+\frac{C}{4\pi} \right) +\frac{R}{16\pi}\left( \frac{1}{G_\ssB}
+\frac{B}{12\pi} \right) +\frac{R^2}{4\pi}\left(\a_\ssB
+\frac{A}{576\pi}\right)\right.\nonumber \\
&&\qquad\qquad\left.
+\frac{1}{4\pi}R_{ab}R^{ab}\left(\b_\ssB-\frac{A}{1440\pi}\right)
+\frac{1}{4\pi}R_{abcd}R^{abcd}\left(\g_\ssB+\frac{A}{1440\pi}\right)
+\ldots\right] \nonumber  \\ \label{effgrav}
\eeqa
where in the action we discard the total derivative term $\Box R$
occurring in $a_2$. In particular from eq.~\rrc{effgrav},
we obtain the renormalized Newton's constant
\beq
\frac{1}{G_\ssR}=\frac{1}{G_\ssB}+\frac{B}{12\pi}\ \ .
\label{Grenorm}
\eeq
In eq.~(\ref{effgrav}), divergent renormalizations also occur for
the cosmological constant $\L_\ssB$ and the quadratic-curvature
coupling constants $\a_\ssB$, $\b_\ssB$ and $\g_\ssB$.
For large values of $\mu$, the constants $A,\ B$ and
$C$ grow to leading order as $\ln(\mu/m),\ \m^2$ and $\m^4$,
respectively, but they also contain subleading and finite
contributions. The higher order bare coupling
constants (beyond those explicitly shown) would receive finite
renormalizations from the finite terms in the one-loop action \rrc{leff},
but they will play no role in the present analysis.

In the following, we will actually consider a \RN\ black hole, because
it provides a more sensitive test of our comparison between the above
results and those in 't Hooft's calculation of black hole entropy.
Thus our background implicitly includes both a metric and a
$U(1)$ gauge potential. Therefore the complete action should be
supplemented with a Maxwell term and, in general, additional higher derivative
interactions with the metric and gauge fields:
\beq
I_{U(1)}=\int d^4x\,\sqrt{-g}\,\left[
- \frac{1}{4} F_{ab}F^{ab} +\delta_\ssB (F_{ab}F^{ab})^2
+\lambda_\ssB R_{ab}\,F^{ac}F_c{}^b+\ldots\right]
\label{uaction}
\eeq
Despite introducing a background gauge field, we consider only
a neutral scalar field as above, and therefore, in the effective action, the
gauge field interactions are completely unaffected by the
scalar one-loop contributions. An obvious extension of the present
analysis would be to repeat the calculations for
a complex scalar field which couples to the gauge potential.

\section{Entropy calculation} \label{entropysect}

In Ref.~\cite{thooft}, 't Hooft attempted to provide a
microphysical explanation
of black hole entropy by considering the modes for a scalar field in the
vicinity of a black hole. In such a calculation, one finds a divergence in
the number of modes because of the infinite blue shift at the event
horizon. To regulate his calculation, 't Hooft introduced a ``brick
wall'' cut-off, demanding that the scalar field vanish within some fixed
distance of the horizon. 't Hooft introduced
this ``simple-minded'' cut-off as an attempt to mimic what he hoped would
be a true physical regulator arising from gravitational interactions.
In the present calculation, we will find that the Pauli-Villars regulator
introduced in the previous section can be used to implement a covariant cut-off
in this entropy calculation, and 't Hooft's brick wall may thereby be removed.
In this way, it is possible to make an explicit comparison of the
divergences appearing in the entropy and in the effective action.

Our calculation follows that of Ref.~\cite{thooft}, but
we consider the more general case of a
\RN\ (RN) black hole, whose metric can be written in the form
\beq
ds^2=- \prmrp    dt^2  + \prmrpb^{-1}   dr^2 +r^2 d\O^2
\label{schwarzschild}
\eeq
where $d\O^2$ is the angular line element for a unit two-sphere.
We assume a non-extremal RN black hole with $r_+>r_-$, so that
$r=r_+$ and $r_-$ correspond to the positions of the outer event
horizon and the inner Cauchy horizon, respectively.
Results for the Schwarzschild black hole are recovered
by simply letting $r_- \rightarrow 0 $.
In this RN background, we consider a minimally coupled neutral scalar field
as in \eqc{miniaction}, which satisfies the Klein-Gordon equation
\beq
(\Box -m^2)\p=0\ .
\label{eomscalar}
\eeq

As described above, 't Hooft's procedure consists of introducing
a brick wall cut-off near the event horizon by setting
\[
\phi(x)=0 \quad {\rm for}\quad r\le {\hrz}+h
\]
with $h\ll{\hrz}$.
To eliminate infrared divergences,
a second cut-off is introduced at a large radius $L\gg {\hrz}$:
$\p(x)=0$ for $r \geq L$.

Expanding the
scalar field in spherical coordinates $\p=e^{iEt}\, Y_{\ell m}(\t,\varphi
)\, f(r)$,
we find that the Klein-Gordon equation becomes
\beq
\frac{r^2E^2}{(r-r_+)(r-r_-)} f(r) +\frac{1}{r^2}
\partial_r[ (r-r_-)(r-{\hrz})\partial_r f(r)]
- \left(\frac{\ell(\ell+1)}{r^2} +m^2 \right) f(r)=0\ .
\label{kgexp} \eeq
In the WKB approximation, one writes $f(r)=\rho(r) e^{iS(r)}$, where $\rho(r)$
is a slowly varying amplitude and $S(r)$ is a rapidly varying phase.
To leading order, only first derivatives of the phase are important.
In particular, eq.~(\ref{kgexp}) yields the radial wave number
$k(r,\ell,E)\equiv\partial_r S$:
\[
k^2=\prmrpb^{-2}\left[
E^2 -\prmrp \left( \frac{\ell(\ell+1)}{r^2}
+m^2 \right)\right]\ .
\]
The number of modes with energy not exceeding $E$ is determined by
summing over the degeneracy of the angular modes, and finding
the radial mode number by counting the number of nodes in the radial
wave function:
\beqa
g(E) &\equiv&\int d\ell \, (2\ell+1) \int_{{\hrz}+h}^{L}dr
\,\frac{1}{\pi}\, k(r,\ell,E) \label{gE} \nonumber\\
&=& \frac{1}{\pi}\int_{{\hrz}+h}^{L} dr \,
 \prmrpb^{-1}\ \ \ \times  \nonumber\\ && \ \ \ \ \ \ \ \ \ \ \  \int d\ell\,
(2\ell+1) \left[ E^2-\prmrp \left(
\frac{\ell(\ell+1)}{r^2} +m^2 \right) \right]^{1/2}\ . \label{twenty}
\eeqa
Above, the sum over the angular quantum number $\ell$ has also been
approximated by an integral, and implicitly this
integration runs over the values of $\ell$ for which the square root
in the integrand is real.

To determine the thermodynamic properties of this system,
we consider the free energy of a thermal ensemble of scalar
particles with an inverse temperature $\b$
\beq
\b F = \sum_N\, \ln\! \left( 1-e^{-\b E_N} \right) \ \ .
\label{barfa}
\eeq
Using eq.~\rrc{twenty}\ to determine the density of states, we obtain
\beqar
\b F& =& \int_0^\infty dE\ \frac{dg}{dE}(E)
\ \ln \left( 1-e^{-\b E} \right) \\
&=& -\frac{\beta}{\pi}
\int_0^\infty  \frac{dE}{e^{\b E}-1}\,\int_{{\hrz}+h}^{L}dr \,
\prmrpb^{-1}\quad\ \times \\
& & \qquad\qquad \int d\ell \, (2\ell+1)\,\left[ E^2-\prmrpb \left(
\frac{\ell(\ell+1)}{r^2}
+m^2 \right) \right]^{1/2}
\eeqar
where an integration by parts has been used to produce the second line.
The integral over $\ell$ can be evaluated to yield
\[
F= -\frac{2}{3\pi} \int_0^\infty
 \frac{dE}{e^{\b E}-1}\, \int_{{\hrz}+h}^{L} dr\,
\prmrpb^{-2} r^2 \left[ E^2 -\prmrp m^2 \right]^{3/2}
\]
where the remaining integration is still taken for values where the square
root is real. The necessity of the brick wall cut-off is clear at this
point, since the integrand diverges with a double pole at the event
horizon, \ie as $r \rightarrow {\hrz}$.
This divergence is more easily examined by introducing a new variable,
$s=1-{\hrz}/r$. The free energy is then given by
\beq
F=-\frac{2  {{\hrz} }^3}{3\pi} \int_0^\infty
\frac{dE}{e^{\b E}-1} \int_{h'}^{L'}
\frac{ds}{s^2 (1-s)^4 (1-u+u s)^2} \left[ E^2-s (1-u+u s) m^2 \right]^{3/2}
\label{free1}
\eeq
where $u=r_-/ r_+$,
$L'=1-{\hrz}/L$ and $h'=h/(\hrz+h)\simeq h/{\hrz}$. Thus, for
small values of $s$, we have $\int_{h'}ds/s^2\simeq-1/h'$,
which diverges as the brick wall is pulled back to the horizon, \ie
as $h'\rightarrow0$.

Now, rather than considering a single scalar field, we repeat 't Hooft's
calculation for the Pauli-Villars regulated field theory introduced in
eq.~\rrc{pvaction}. Each of the fields makes a contribution to the
free energy as in \eqc{free1}, and the total free energy
becomes
\beq
\bar{F}=-\frac{2 {\hrz}^3}{3\pi}\,\sum_{i=0}^5\Delta_i
\, \int_0^\infty
\frac{dE}{e^{\b E}-1} \int_{h'}^{L'}
\frac{ds}{s^2 (1-s)^4 (1-u+u s)^2 } \left[ E^2-s (1-u+u s) m_i^2 \right]^{3/2}
\label{freepv}
\eeq
where $\Delta_0=\Delta_3=\Delta_4=+1$ for the commuting fields,
and $\Delta_1=\Delta_2=\Delta_5=-1$ for the anticommuting fields.
The free energy of the anticommuting regulator fields comes with
a minus sign with respect to the commuting fields, as is required since
the role of these fields is to cancel contributions of very high energy
modes in the regulated theory. Now, if we examine
the divergence of the revised free energy in
\eqc{freepv}, which arises for small $s$,
we find $\sum_{i=0}^5\Delta_i\int_{h'}ds/s^2
=0$; there is a precise cancellation between the original scalar and the
regulator fields. Similarly, we find that
a sub-leading logarithmic divergence at small $s$ is also
cancelled, since $\sum_{i=0}^5\Delta_im_i^2=0$.

Thus, in the Pauli-Villars regulated theory, we are free
to remove 't Hooft's brick wall\footnote{Note that our total free energy is
actually defined
through the limit $h'\rightarrow0$, where the brick wall still played a
role in defining the density of states for the individual fields.
We assume that the results from this limiting procedure coincide with those
arising within the canonical quantization of the Pauli-Villars
regulated theory.}. Setting $h'=0$, our
expression for the free energy becomes
\beq
\bar{F}=-\frac{2{\hrz}^3}{3\pi}\, \int_0^\infty
\frac{dE}{e^{\b E}-1} \int_{0}^{L'}
\frac{ds}{s^2 (1-s)^4 (1-u+u s)^2 } \,
\sum_{i=0}^5\Delta_i\left[ E^2-s (1-u+u s) m_i^2
\right]^{3/2}\ \ .
\label{freepv2}
\eeq
Now, integrating over $s$ and $E$, we focus only on the divergent
contributions at the horizon and find
\beq
\bar{F}\simeq
-{\hrz}^3\left[ \frac{ \pi}{6(1-u) \b^2 }B+\frac{4 \pi^3 (2-3u)}
{45 (1-u)^3\b^4} A
\right]
\label{barf}
\eeq
where $A$ and $B$ are the same constants given in
eqs.~\rrc{Acoeff}\ and \rrc{Bcoeff}, respectively.
We emphasize that eq.~(\ref{barf}) neglects contributions to the integral
which do not diverge as $\m\rightarrow \infty$.
The entropy is then given by
\beq
S=\b^2  \frac{\partial\bar{F}}{\partial \b}
= {\hrz}^3 \left[ \frac{ \pi}{3  (1-u) \b} B +\frac{16 (2-3u) \pi^3}
{45 (1-u)^3 \b^3} A
\right] \ \ .  \label{entro}
\eeq
Choosing the inverse temperature $\beta$ to correspond to
the Hawking temperature of a non-extremal RN black hole, we set
\[
\b= \frac{4\pi{\hrz}}{1-u}\ ,
\]
upon which the   entropy  (\ref{entro}) becomes
\beq
S=\frac{\A}{4} \frac{B}{12\pi}  +\frac{(2-3u) A}{180}
\label{entrofin}
\eeq
where $\A=4\pi {\hrz}^2$ is the surface area of the event horizon.
Thus we see that
the entropy contains   the constants $A$ and $B$, which give
precisely the dependence on the regulator mass $\mu$ appearing
in the renormalization of Newton's constant and of the quadratic-curvature
coupling constants.

\section{Discussion} \label{concsect}

\subsection{Renormalization of the Entropy}

The entropy \rrc{entrofin}\ calculated in
section 3, and the standard Bekenstein-Hawking entropy, \ie
$S_{\ssc BH}=\A/(4G)$, are related in a simple way.
If the latter is written in terms of the
bare Newton's constant, then adding these two entropies yields
\beqa
S_{\ssc BH}+S&=&\frac{\A}{4}\left(\frac{1}{G_\ssB}+\frac{B}{12\pi}\right)
+\frac{(2-3u) A}{180} \nonumber \\
&=&\frac{\A}{4G_\ssR}+\frac{(2-3u)A}{180}
\label{pasum}
\eeqa
where we have used \eqr{Grenorm} for the renormalized Newton's
constant. Hence we find
that the first contribution proportional to $B$ in the scalar field
entropy provides precisely the one-loop renormalization of the
Bekenstein-Hawking entropy. Thus these terms combine in precisely
the manner suggested by Susskind and Uglum\cite{susskind}.

Note that we must still account for the constant term
proportional to $A$ which appears in \eqc{entrofin}.
Following the recent work on black hole entropy for
higher curvature effective gravitational
actions[4--8], 
we expect that this constant contribution to the entropy should be related to
the quadratic-curvature interactions in the action \rrc{actg}. In particular,
we must consider
\[
I_{2}=
\int d^4x\, \sqrt{-g} \left[
\frac{\a_\ssB}{4\pi} R^2 +
\frac{\b_\ssB}{4\pi} R_{ab}R^{ab} +
\frac{\g_\ssB}{4\pi} R_{abcd}R^{abcd} \right]\ \ .
\]
These interactions will modify the standard result for black hole entropy,
\ie the Bekenstein-Hawking entropy, by adding an expression of the form
\beq
\Delta S=\oint d^2x\sqrt{h}\left[
2\a_\ssB R + \b_\ssB g^{ab}_\bot R_{ab} - \g_\ssB R^{abcd}\hat\varepsilon_{ab}
\hat\varepsilon_{cd} \right]\ \ .
\label{quads}
\eeq
Here,
the integral is evaluated over a space-like cross section of
the event horizon, $g^{ab}_\bot$ is the
metric in the normal subspace to this cross section, and
$\hat\varepsilon_{ab}$ is the binormal to the cross section
--- for more details see either Ref.~\cite{tedtwo} or \cite{vivek}.
When eq.~(\ref{quads}) is evaluated in the present background,
only the last two terms  make  a nonvanishing
contribution (since $R=0$ for a four-dimensional RN black hole):
$\Delta S= -8 \pi u \b_\ssB +16\pi (1-2u)\g_\ssB$.
Now, including this contribution
along with the previous terms in   (\ref{pasum}), we  obtain the total
black hole entropy
\beqa
S_{\rm total} &=&   S_{\ssc BH}+\Delta S+S  \\ \nonumber&=&
\frac{\A}{4}\left(\frac{1}{G_\ssB}+\frac{B}{12\pi}\right)
-8\pi u (\b_\ssB-\frac{A}{1440\pi}) +
16\pi (1-2 u)\left(\g_\ssB +\frac{A}{1440\pi}\right)\\
&=&\frac{\A}{4G_\ssR} -8\pi u \b_\ssR +16\pi(1-2u)\g_\ssR
  \label{totals}
\eeqa
where, as in \eqc{effgrav}, we have the renormalized coupling constants:
 $\b_\ssR=\b_\ssB -\frac{A}{1440\pi}$ and
$\g_\ssR=\g_\ssB +\frac{A}{1440\pi}$. Thus both terms in the scalar field
entropy \rrc{entrofin}\ account for
precisely the scalar one-loop renormalization of the
full black hole entropy.
Note that for a Schwarzschild black hole (\ie that which arises with $u=0$),
the contribution of $\b_\ssR$ in eq.~(\ref{totals}) vanishes
because the background curvature would satisfy $R_{ab}=0$. Thus, choosing
a RN background allows for a more sensitive comparison between the
renormalization of the effective action and 't Hooft's entropy
calculation. The appearance of subleading terms
in the scalar field entropy (\ref{entrofin}), and their interpretation
in terms of higher
curvature contributions to the black hole entropy, have also been discussed
  in an
alternate field-theoretic calculation of black hole entropy in
Ref.~\cite{solodukhin}.

A priori, one might not have expected
the Pauli-Villars scheme to regulate
't Hooft's entropy calculation at all. In fact, though, not only do we find
that the Pauli-Villars scheme regulates the latter calculation,
our results are in complete agreement
with the suggestion of Susskind and Uglum. The divergences appearing
in 't Hooft's statistical-mechanical calculation of black hole entropy are
precisely the quantum field theory divergences associated with the
renormalization of the coupling constants appearing in the expressions
of the entropy. This identification includes
both the divergent
and finite contributions in the renormalization of the couplings,
$G_\ssB$, $\b_\ssB$ and $\g_\ssB$.
This precise equality, including the finite
terms, occurs because the combinations
of masses $\sum\Delta_im^2_i\ln m^2_i$ and $\sum\Delta_i\ln m^2_i$
arise naturally in both calculations. We have not considered here
any of the remaining
finite contributions arising in the free energy \rrc{freepv2}. It
should be possible to identify the corresponding contributions to the
black hole entropy with finite renormalization of the
higher curvature terms arising from finite terms in the one-loop action
\rrc{leff}. There is also a class of contributions
to the free energy depending on the infrared cut-off. (Of course,
these terms cannot be avoided with the Pauli-Villars regulator, which
is an ultraviolet regulator.) To leading order,
these infrared terms yield the usual (extensive) free energy for a gas
of free scalar particles enclosed in a volume $\frac{4}{3}\pi L^3$.
There are also lower order terms, which arise due to the curved
space-time geometry.

\subsection{Extremal Reissner-Nordstrom}

It is not difficult to repeat our calculations for the
case of an extremal RN black hole with $r_ +=r_-$. In this case,
't Hooft's brick wall cut-off leads to ill-defined results\cite{extreme2}.
The problem is that the coordinate cut-off, $h$, cannot be converted to
a proper length cut-off because any point which is a fixed coordinate
distance outside of the extremal horizon is in fact an infinite proper
distance from the horizon (on a constant time hypersurface).
No such problem arises with the covariant Pauli-Villars regulator.
However, precisely at the extremal
limit $u=1$, the structure of the small $s$ divergences in
eq.~(\ref{freepv2}) changes, and hence we must re-evaluate the
integral. We find that the divergent part of the free energy
is given by
\[
\bar{F}_{ext}\simeq-{\hrz}^3\left[ \frac{ \pi}{3  \b^2 }B+\frac{4 \pi^3  }
{ 9 \b^4} A
\right] \ ,
\]
and the entropy which follows is
\beq
 S_{ext}= {\hrz}^3\left[ \frac{2 \pi}{3  \b  }B+\frac{16 \pi^3  }
{ 9 \b^3} A
\right] \ . \label{sext}
\eeq
Here $A$ and $B$ are the same divergent coefficients (\ref{Acoeff})
and (\ref{Bcoeff}) that appear in the scalar one-loop action and
in the non-extremal entropy. Hence, with a covariant regulator,
we find that the extremal entropy has no stronger divergences
than appear in the non-extremal case. In fact, the entire result
has essentially the same form as the non-extremal entropy in
\eqc{entro}.

To proceed further,  one must fix the inverse temperature
in \eqc{sext}. Using the standard formula\cite{hawking}, $T=\kappa/(2\pi)$
where $\kappa$ is the surface gravity, one finds that the
temperature is zero since the surface gravity vanishes for the
extremal RN black hole. Thus, the inverse temperature
$\beta$ diverges, and we find that $S_{ext}$ vanishes\footnote{The
same is true when using the brick wall regulator\cite{extreme2}.}.
This result is
in accord with the recent discovery\cite{extreme} that extremal
black holes should have vanishing entropy, since one expects then that
the renormalization contribution must also vanish; since the
value of entropy is independent of the coupling constants, the renormalized
value of zero is still zero. Note that the integral for the
free energy in \eqc{freepv2}\ is finite for all $u$ in $0\le u\le1$.
Hence, if one could evaluate the full regulated free energy, the resulting
entropy would exhibit a smooth transition from the behavior appearing in
\eqc{entrofin} for $u<1$
to zero at $u=1$. Further the transition should occur for $u-1=O(m/\mu)$.
Thus we speculate that the total black hole entropy for
which we are calculating the one-loop renormalization should also
make a smooth transition to $S=0$ in the extremal limit, but that
the precise details of the transition would depend on the
ultraviolet characteristics of the underlying theory of quantum
gravity.

On the other hand,
the recent investigations of extremal black holes\cite{extreme}
also suggest that an extremal black hole can be in equilibrium with
a heat bath of an arbitrary temperature. Hence, one might consider
leaving the inverse temperature arbitrary in \eqc{sext}. In this
case, one has the curious result that $S_{ext}$ appears to represent
the renormalization of some finite entropy expression for
an extremal RN black hole. For example, the first term in
\eqc{sext} would represent the renormalization of
$S={\A\over4G}{{8\pi r_+}\over\beta}$. Previous calculations have given no
indication that such an entropy arises for extremal black holes, and
so one may conclude that one must use $\beta\rightarrow\infty$ in
this case. Alternatively, it may be that 't Hooft's model does
not capture the full physics of extremal black holes, and that the
correct result should still be $S_{ext}=0$ even with a nonvanishing
temperature.

\subsection{On-shell/Off-shell}

With few exceptions\cite{solodukhin,except}, discussions and derivations
of black hole entropy
focus on black hole backgrounds which are saddle-points of
the gravitational action under investigation, \ie backgrounds
which are solutions of the equations of motion (see, for example,
Refs.~\cite{ted,wald,visser}). We will now discuss this point in the
context of the present calculation. Our RN background, which includes
both the metric of \eqc{schwarzschild} and, implicitly,
a vector potential $A=Q/(4\pi r)\,dt$,
solves the Einstein-Maxwell equations with $4\pi r_+r_-=G Q^2$.
Demanding that the background be a solution of the equations
arising from the total effective action
would require that the renormalized cosmological constant vanish,
that (most of)
the renormalized coupling constants for the higher derivative interactions
vanish\footnote{Since $R=0$ for the
four-dimensional RN background, the metric would still
solve the higher derivative equations of motion if $\a_\ssR$ was
nonvanishing. Further, because the combination
$R_{abcd}R^{abcd}-4R_{ab}R^{ab}+R^2$ is a topological density
in four dimensions
which does not affect the equations of motion, $\b_\ssR=-4\g_\ssR$
would ensure that the RN black hole solves the renormalized equations
of motion. Generically, though, one expects for the higher
derivative interactions in the effective action (including those
not explicitly shown in eqs. \rrc{effgrav}\ and \rrc{uaction})
that the corresponding coupling constant must be set to zero
to ensure that the RN background is a solution.},
and that the renormalized Newton's constant appear
in the above equation with $r_+$, $r_-$ and $Q$.

For the present calculations, first of all we note that
this question of whether or
not the background solved any equations of motion is
irrelevant for 't Hooft's entropy calculations in section 3.
Further, note that if the RN black hole is a solution of the renormalized
equations of motion, it cannot at the same time solve the bare
equations of motion (which may appear more appropriate for section 3).
In our discussion, though,
we use the usual entropy expressions (\ie the Bekenstein-Hawking
formula, and the higher curvature corrections in \eqc{quads}) to
assign our (single) background a black hole entropy within
both the renormalized and the bare theories. Thus the present
calculations treat these entropy formulas as being valid off-shell,
\ie valid for backgrounds which do not satisfy the equations of motion.
This possibility is suggested in Ref.~\cite{solodukhin},
which presented a derivation of black hole entropy which made no
explicit use of the equations of motion.

\subsection{Robustness}

One would like to know whether the present results hold
for arbitrary field theories coupled to gravity, rather than for
just a minimally coupled scalar field. One simple extension
of our calculations would be to consider a non-minimally coupled
scalar field. The original matter action in
eq.~\rrc{miniaction}\ is then extended to
\[
I'_m=-\frac{1}{2} \int d^4x \, \sqrt{-g} \left[ g^{a b} \nabla_{\! a}\phi
 \nabla_{\! b} \phi +m^2 \phi^2 +\xi R\phi^2\right]\ .
\]
It is well known\cite{birrell} that the additional coupling of
the scalar field to the curvature
modifies the adiabatic expansion coefficients in \eqc{expcoeff}, and
therefore it affects the renormalizations of the bare coupling constants.
For example, \eqc{Grenorm} for the renormalized Newton's constant
is replaced by
\beq
\frac{1}{G_\ssR}=\frac{1}{G_\ssB}+\frac{B}{2\pi}\left(\frac{1}{6}-\xi
\right)\ \ .
\label{newG}
\eeq
On the other hand, if we repeat the calculation of section 3 for the
new scalar field theory, we find that the resulting entropy is
completely unchanged. The new coupling constant
$\xi$ enters the new equation of motion, $(\Box -m^2-\xi R)\p=0$,
which replaces \eqc{eomscalar}.
The remainder of the calculation is unmodified, though, because $R=0$ for
the background RN metric. Given that Newton's constant is renormalized
as in \eqc{newG},
the entropy in eq.~\rrc{entrofin}, which is independent of $\xi$,
does not properly account for
the renormalization of the Bekenstein-Hawking formula.

We see this failure as a limitation of 't Hooft's model for the
calculation of black hole entropy. It is clear that this model
does not capture the full physics of the problem. For example,
the free energy in \eqc{barf}\ does not contain a quartic divergence
as would be expected from the renormalization of the cosmological
constant in \eqc{effgrav}. This omission can be traced to the fact that
\eqc{barfa}\ does not include a contribution from
the zero point energies. Of course, neglecting this contribution
is entirely appropriate for a leading order WKB calculation.

The Euclidean path integral would provide an alternate approach
to this calculation\footnote{At this point, we note that in introducing
the regulator
fields in \eqc{freepv}, it may appear that choosing $\Delta_i=-1$
and a Bose-Einstein distribution (rather than a Fermi-Dirac
distribution) for the anticommuting fields is somewhat arbitrary.
In the Euclidean path integral,
though, both of these choices are completely fixed.}.
In fact, this approach provides a more rigorous framework
to study black hole entropy.
We are presently adapting our analysis to the
Euclidean path integral, and expect that it will yield the
correct renormalization of black hole entropy, even for non-minimally
coupled scalar fields.

\section*{Acknowledgments}
This research was supported by NSERC of Canada and Fonds FCAR du
Qu\'ebec. We would like to thank Cliff Burgess,
Keith Dienes, Noureddine Hambli, Ted Jacobson
and Alex Marini for useful conversations. RCM would also
like to thank the Institute for Advanced Study for its hospitality
while a portion of this paper was written.

\end{document}